\documentclass[11pt]{article}
\usepackage{amssymb}

\usepackage[totalwidth=460truept,totalheight=600truept]{geometry}
\usepackage{latexsym,amssymb,amsmath,graphicx,accents,eucal,slashed,subfigure,hyperref}
\usepackage{dsfont}
\usepackage[T1]{fontenc}

\global\arraycolsep=1truept

\numberwithin{equation}{section}

\begin{document}

\bigskip \hfill IFUP-TH 2012/21

\vskip 1.4truecm

\begin{center}
{\huge \textbf{Renormalization Of Gauge Theories}}

\vskip .5truecm

{\huge \textbf{Without Cohomology}}

\vskip 1truecm

\textsl{Damiano Anselmi}

\vskip .2truecm

\textit{Dipartimento di Fisica ``Enrico Fermi'', Universit\`{a} di Pisa, }

\textit{and INFN, Sezione di Pisa,}

\textit{Largo B. Pontecorvo 3, I-56127 Pisa, Italy}

\vskip .2truecm

damiano.anselmi@df.unipi.it

\vskip 1.5truecm

\textbf{Abstract}
\end{center}

\medskip

{\small We investigate the renormalization of gauge theories without
assuming cohomological properties. We define a renormalization algorithm
that preserves the Batalin-Vilkovisky master equation at each step and
automatically extends the classical action till it contains sufficiently
many independent parameters to reabsorb all divergences into
parameter-redefinitions and canonical transformations. The construction is
then generalized to the master functional and the field-covariant proper
formalism for gauge theories. Our results hold in all manifestly
anomaly-free gauge theories, power-counting renormalizable or not. The
extension algorithm allows us to solve a quadratic problem, such as finding
a sufficiently general solution of the master equation, even when it is not
possible to reduce it to a linear (cohomological) problem.}

\vskip 1truecm

\vfill\eject

\section{Introduction}

\label{RGren} \setcounter{equation}{0}

When a gauge theory is power-counting renormalizable, studying its
renormalization is not a difficult task, because the terms contained in the
classical action and the counterterms are finitely many, and usually a small
number, and it is possible to write down all of them and work out their
transformation properties explicitly. When, however, composite fields of
higher dimensions are turned on, infinitely many others must be included. If
the theory is not power-counting renormalizable, like quantum gravity, we
have to deal with infinitely many terms in any case. In these situations,
cohomological properties can simplify several tasks, because they allow us
to classify terms and counterterms into gauge-invariant ones, gauge-trivial
ones, and gauge non-invariant ones. This classification is useful to prove
that divergences can be subtracted redefining the ingredients of the
classical action, that is to say parameters, fields and sources, or prove
that it is possible to extend the classical action so that divergences can
be subtracted that way. A natural question arises whether cohomological
properties are essential for the renormalization of gauge theories or not.
In this paper we prove that they are not.

The only assumption we make is that the theory is manifestly free of gauge
anomalies (global symmetries are instead allowed to be anomalous), which
means that there must exist a regularization such that the classical action $%
S_{c}$ satisfies the Batalin-Vilkovisky master equation $(S_{c},S_{c})=0$ 
\cite{bata} exactly at the regularized level. Without using or assuming
cohomological properties we show that it is always possible to extend $S_{c}$%
, preserving the master equation, till the extended $S_{c}$ contains enough
independent parameters to subtract all divergences by means of
parameter-redefinitions and canonical transformations. A classical action
with these properties is called \textit{parameter-complete}. In our approach
it is renormalization itself that guides us through the appropriate $S_{c}$%
-extensions, till parameter-completion is achieved. Among the other things,
parameter-completion is necessary to have renormalization-group (RG)
invariance.

We do not assume that the gauge algerba closes off shell, nor that the
number of independent parameters necessary to renormalize divergences is
finite. The search for theories that are renormalizable with a finite number
of independent parameters, and do not obey known power-counting criteria, is
out of the purposes of this paper. Nevertheless, we do believe that the
formalism developed here will help organize that search in a more convenient
way.

We can illustrate the basic idea of our extension algorithm in a simple case
that involves no gauge symmetries. Consider the massless $\varphi ^{6}$%
-theory 
\[
S_{\text{scal}}(\varphi ,\lambda _{6}\mu ^{2\varepsilon })=\frac{1}{2}\int
(\partial _{\mu }\varphi )(\partial ^{\mu }\varphi )-\lambda _{6}\mu
^{2\varepsilon }\int \frac{\varphi ^{6}}{6!}, 
\]
in four dimensions, using the dimensional-regularization technique. The
coupling $\lambda _{6}$ has dimension $-2$ and $\varepsilon =4-D$, $D$ being
the continued dimension. RG invariance is apparent if we hide the parameter $%
\mu $ inside the bare coupling $\lambda _{6\mathrm{B}}=\lambda _{6}\mu
^{2\varepsilon }$. Now we calculate the one-loop divergences and subtract
them just as they come. In the minimal subtraction scheme the one-loop
renormalized action reads 
\begin{equation}
S_{\text{scal}}^{\text{1-loop}}(\varphi ,\lambda _{6}\mu ^{2\varepsilon
},\mu ^{-\varepsilon }\hbar /\varepsilon )=S_{\text{scal}}-\frac{35\hbar
\lambda _{6}^{2}\mu ^{3\varepsilon }}{(4\pi )^{2}\varepsilon }\int \frac{%
\varphi ^{8}}{8!},  \label{olg}
\end{equation}
which is clearly not RG invariant. The reason is that it misses an
independent parameter for the new vertex $\varphi ^{8}$. However, it does
depend on a new quantity, which is $\mu ^{-\varepsilon }\hbar /\varepsilon $%
. Therefore it is sufficient to replace $\hbar /\varepsilon $ with a new
dimensionless parameter $\lambda _{8}^{\prime }$, and define the extended RG
invariant classical action 
\begin{equation}
S_{\text{scal}}^{(1)}(\varphi ,m,\lambda _{6}\mu ^{2\varepsilon },\lambda
_{8}^{\prime }\mu ^{-\varepsilon })=\frac{1}{2}\int (\partial _{\mu }\varphi
)(\partial ^{\mu }\varphi )-\lambda _{6}\mu ^{2\varepsilon }\int \frac{%
\varphi ^{6}}{6!}-\frac{35\lambda _{6}^{2}\lambda _{8}^{\prime }\mu
^{3\varepsilon }}{(4\pi )^{2}}\int \frac{\varphi ^{8}}{8!},  \label{olg2}
\end{equation}
with $\lambda _{8\mathrm{B}}^{\prime }=\lambda _{8}^{\prime }\mu
^{-\varepsilon }$ at the tree level. The divergence contained in (\ref{olg})
can now be reabsorbed into a renormalization of $\lambda _{8}^{\prime }$,
which reads at one loop 
\begin{equation}
\lambda _{8\mathrm{B}}^{\prime }=\mu ^{-\varepsilon }\left( \lambda
_{8}^{\prime }+\frac{\hbar }{\varepsilon }\right) .  \label{run}
\end{equation}
This is not the end of the story, however, not even at one loop. For
example, the classical action $S_{\text{scal}}^{(1)}$ generates one-loop
counterterms $\sim \varphi ^{10}$. Thus we need to iterate the procedure,
introduce a new parameter $\lambda _{10}^{\prime }$, and proceed like this
indefinitely.

In this simple example what we have done is redundant. On the other hand,
when gauge symmetries are present and cohomological theorems do not help us,
extending the classical action preserving the master equation is not so
straightforward. Nevertheless, we can use the strategy just sketched and let
renormalization build the extended classical action by itself. Observe that
the parametrization we obtain from this kind of procedure is a bit unusual.
Indeed, the vertex $\varphi ^{8}$ in (\ref{olg2}) in not just multiplied by
some parameter $\lambda _{8}$, but by a complicated product of parameters.
Here we can just replace the coefficient of $\varphi ^{8}/8!$ with $-\lambda
_{8}\mu ^{3\varepsilon }$, but in the most general case we cannot assume
that a nice parametrization exists. This forces us to work with the unusual
one. The perturbative expansion is also organized in an unusual, but
consistent way. If take $\lambda _{6}\sim g^{4}$ and $\lambda _{8}^{\prime
}\sim 1/g^{2}$, with $g\ll 1$, the parameter $\lambda _{8}^{\prime }$ is
large, but it always appears inside combinations that are altogether small.
The running predicted by (\ref{run}) is also consistent, since $\beta
_{8}^{\prime }=\hbar $ is turned into $\beta _{g}\sim \hbar g^{3}$.

Switching to gauge theories, the results of this paper prove that all
divergences generated by renormalization fit into suitable gauge-invariant
extensions of the action. The price is that the gauge symmetry itself may be
extended, or modified in a non-trivial way, because there is no guarantee
that after the extension procedure the final gauge symmetry will be
equivalent to the starting one. At the same time, our results do not prove
that all gauge-invariant terms we can construct fit into extensions of the
action. In other words, if we can construct some gauge-invariant terms that
do not fit into an extended action, renormalization will never be able to
generate them as counterterms. Furthermore, the parameter-complete action we
obtain may not be the most general extended action. It is just the minimally
extended action that can reabsorb all divergences into
parameter-redefinitions and canonical transformations.

We emphasize the main issue, here. The master equation $(S_{c},S_{c})=0$ is
quadratic in the action $S_{c}$, thus the compatibility of gauge symmetry
and renormalization is encoded in a quadratic problem. Instead,
cohomological problems are linear in $S_{c}$. When we assume cohomological
properties we in practice assume that the quadratic problem can be reduced
to a linear one, which makes life much easier. It is nice to know that if
cohomological properties do not hold, or are not assumed to hold, we can
still build the action we need, even if the problem cannot be linearized and
at every step the subtraction algorithm becomes more and more involved. The
procedure we outline is conceptually simple, but rather involved at the
practical level. At this stage, its most important applications appear to be
theoretical.

Cohomological properties provide a purely algebraic classification and have
no strict relation with the renormalization algorithm. Typically, they
ensure that all gauge-invariant terms we can construct, even those that
renormalization cannot generate as counterterms, can be included extending
the classical action. From the algebraic point of view, cohomological
theorems can be more general than our results. That kind of generality,
however, may be unnecessary for the purposes of renormalization. At the same
time, our construction is more general in a different direction, because it
also works when cohomological theorems do not hold or are unavailable.

The classical action $S_{c}$ we start from can be any particular local
solution of the master equation. Then we use the properties of
renormalization to build the parameter-complete local extension $S_{\subset
} $. The extension map 
\begin{equation}
S_{c}\rightarrow S_{\subset }  \label{extamap}
\end{equation}
is also a powerful machine to prove the existence of new solutions of the
master equation, even when we are unable to write them down explicitly.

In the last part of the paper we generalize our results to the master
functional defined in ref. \cite{mastercan} and the field-covariant \textit{%
proper formalism} for gauge theories. The master functional $\Omega $
satisfies the proper master equation $\lfloor \Omega ,\Omega \rfloor =0$,
where the squared antiparentheses are obtained generalizing the
Batalin-Vilkovisky antiparentheses to the sector made of composite fields
and their gauge transformations. The proper formalism allows us to express
all local perturbative field redefinitions and changes of gauge-fixing as
``proper'' canonical transformations (see section 6 for details), and
interpret them as true changes of variables in the functional integral,
instead of simple replacements of integrands.

The generalization of the results of this paper to theories that are not
manifestly free of gauge anomalies is left to a separate investigation.

Throughout this paper we use the dimensional-regularization technique and
the minimal subtraction scheme. Nevertheless, once the action is extended to
contain enough independent parameters, it is possible to switch to an
arbitrary scheme making finite redefinitions of those parameters. 
From now on we switch to the Euclidean notation.

The paper is organized as follows. In section 2 we briefly recall how the
renormalization algorithm works when cohomological properties hold or are
assumed to hold. In section 3 we formulate the completion algorithm induced
by renormalization. At this stage, we introduce redundant parameters so that
all divergences can be subtracted by means of parameter-redefinitions,
without involving canonical transformations. In section 4 we study the
perturbative expansion and discuss consistent truncations that allow us to
work with the desired precision with a finite number of terms and a finite
number of operations. In section 5 we extend the completion algorithm to
include both canonical transformations and parameter-redefinitions. In
section 6 we generalize the construction to the master functional and the
field-covariant proper formalism for gauge theories. In section 7 we collect
a few remarks to point out some interesting features of our construction. In
section 8 we comment on the search for the most general solution of the
master equation, obtained extending the starting classical action. We show
that in general this strategy does not allow us to achieve our goals.
Section 9 contains our conclusions.

Before starting our investigation we recall a few definitions and facts that
will be useful throughout the paper. The antiparentheses of two functionals $%
X$ and $Y$ of the fields $\Phi $ and the sources $K$ coupled to the $\Phi $%
-gauge transformations are 
\[
(X,Y)=\int \left( \frac{\delta _{r}X}{\delta \Phi ^{A}}\frac{\delta _{l}Y}{%
\delta K_{A}}-\frac{\delta _{r}X}{\delta K_{A}}\frac{\delta _{l}Y}{\delta
\Phi ^{A}}\right) . 
\]
If $S$ denotes the classical action, the generating functionals $Z$ and $W$
are defined by 
\begin{equation}
Z(J,K)=\int [\mathrm{d}\Phi \hspace{0.01in}]\exp \left( -S(\Phi ,K)+\int
\Phi ^{A}J_{A}\right) =\exp W(J,K),  \label{funto}
\end{equation}
while $\Gamma (\Phi ,K)$ is the Legendre transform of $W$ with respect to $J$%
. A general theorem says that $\Gamma $ satisfies the identity 
\begin{equation}
(\Gamma ,\Gamma )=\langle (S,S)\rangle .  \label{useidu}
\end{equation}
This formula can be proved making a change of variables $\Phi \rightarrow
\Phi +\xi (S,\Phi )$ in the functional integral (\ref{funto}), where $\xi $
is a constant anticommuting parameter. In particular, $(S,S)=0$ implies $%
(\Gamma ,\Gamma )=0$. Finally, in dimensional regularization the functional
integration measure $[\mathrm{d}\Phi \hspace{0.01in}]$ is invariant under
arbitrary perturbative changes of field variables.

\section{Renormalization with cohomology}

\setcounter{equation}{0}

In this section we briefly review how renormalization proceeds when suitable
cohomological properties hold, specified in formulas (\ref{chopr}) and (\ref
{autoassu}) below. To make the presentation simpler, here we also assume
that the gauge algebra closes off shell, so there exists a choice of
variables such that $S_{c}$ has the form 
\begin{equation}
S_{c}(\Phi ,K)=\mathcal{S}(\Phi )-\int R^{A}(\Phi )K_{A},  \label{basa}
\end{equation}
where $R^{A}(\Phi )$ is the symmetry transformation of the field $\Phi ^{A}$
. As said, we use the dimensional-regularization technique and the minimal
subtraction scheme, and the starting classical action $S_{c}$ must satisfy
the master equation $(S_{c},S_{c})=0$ exactly at the regularized level.
Among the other things, minimal subtraction scheme means that $S_{c}$ does
not contain evanescent terms equal to the product of finite local terms
times some powers of $\varepsilon =4-D$. Indeed, if terms of this type were
present we would not be able to extract the divergent parts (of master
equations, see below) in an efficient way, since finite local contributions
could originate from products between divergent and evanescent terms.

For example, in pure non-Abelian Yang-Mills theory we have $\Phi
^{A}=(A_{\mu }^{a},C^{a},\bar{C}^{a},B^{a})$, where $A_{\mu }^{a}$ are the
gauge fields, $C^{a}$ and $\bar{C}^{a}$ are the Fadeev-Popov ghosts and
antighosts, respectively, and $B^{a}$ are the Lagrange multipliers for the
gauge-fixing. We write the sources as $K_{A}=(K_{a}^{\mu },K_{C}^{a},K_{\bar{%
C}}^{a},K_{B}^{a})$. The classical action is 
\begin{eqnarray*}
S_{c}(\Phi ,K) &=&\int \left( \frac{1}{4}F_{\mu \nu }^{a\ 2}-\frac{\lambda 
}{2}(B^{a})^{2}+B^{a}\partial \cdot A^{a}-\bar{C}^{a}\partial _{\mu }D_{\mu
}C^{a}\right)  \\
&&-\int D_{\mu }C^{a}K_{\mu }^{a}+\frac{g}{2}\int
f^{abc}C^{b}C^{c}K_{C}^{a}-\int B^{a}K_{\bar{C}}^{a}{,}
\end{eqnarray*}
where $F_{\mu \nu }^{a}=\partial _{\mu }A_{\nu }^{a}-\partial _{\nu }A_{\mu
}^{a}+gf^{abc}A_{\mu }^{b}A_{\nu }^{c}$ is the field strength, $D_{\mu
}C^{a}=\partial _{\mu }C^{a}+gf^{abc}A_{\mu }^{b}C^{c}$ is the covariant
derivative of the ghosts and $f^{abc}$ are the structure constants of the
Lie algebra. The theory is power-counting renormalizable, hence its
renormalization is straightforward. However, we can imagine to consider a
more general class of theories, obtained adding gauge invariant composite
fields of arbitrary dimensions, such as $(F_{\mu \nu }^{a\ 2})^{n}$ with $n>1
$, multiplied by new coupling costants. Then to renormalize the theory we
can either apply the method recalled in this section, which uses
cohomological properties, or follow the strategy developed in the next
sections.

Call $S_{n}$ the action renormalized up to $n$ loops, with $S_{0}=S_{c}$. By
assumption, $S_{n}$ has the form $S_{c}+$poles in $\varepsilon $. Assume, by
induction, that $S_{n}$ also satisfies the master equation $(S_{n},S_{n})=0$
exactly at the regularized level. Then (\ref{useidu}) tells us that the $n$%
-loop renormalized $\Gamma $-functional $\Gamma _{n}$ satisfies the master
equation $(\Gamma _{n},\Gamma _{n})=0$. Call $\Gamma _{n\hspace{0.01in}\text{%
div}}^{(n+1)}$ the $(n+1)$-loop divergent part of $\Gamma _{n}$. By the
theorem of locality of counterterms, $\Gamma _{n\hspace{0.01in}\text{div}%
}^{(n+1)}$ is a local functional. Then the $(n+1)$-loop divergent part of
the master equation $(\Gamma _{n},\Gamma _{n})=0$ gives the cohomological
problem 
\begin{equation}
(S_{c},\Gamma _{n\hspace{0.01in}\text{div}}^{(n+1)})=0.  \label{chopr}
\end{equation}

The cohomological assumption we make now is that the most general solution
of this problem has the form 
\begin{equation}
\Gamma _{n\hspace{0.01in}\text{div}}^{(n+1)}(\Phi ,K)=\mathcal{G}(\Phi
)+(S_{c},\chi ),  \label{autoassu}
\end{equation}
where $\mathcal{G}(\Phi )$ and $\chi (\Phi ,K)$ are local functionals. The
main meaning of (\ref{autoassu}) is that the possibly non-trivial part $%
\mathcal{G}$ depends only on the fields $\Phi $. Theorems that ensure (\ref
{autoassu}) have been proved both for Yang-Mills theory and gravity, for
local composite fields and local functionals of arbitrary ghost numbers \cite
{coho}.

Often we can characterize $\mathcal{G}(\Phi )$ even more precisely. Assuming
that the set of fields $\Phi ^{A}$ is made of the physical fields $\phi $,
the ghosts $C$, plus the gauge-trivial subsystem $\bar{C}$-$B$ made of
antighosts $\bar{C}$ and Lagrange multipliers $B$, then it is possible to
further decompose $\mathcal{G}(\Phi )$ as 
\begin{equation}
\mathcal{G}(\Phi )=\mathcal{G}^{\prime }(\phi )+(S_{c},\chi ^{\prime }),
\label{galassu}
\end{equation}
where $\mathcal{G}^{\prime }(\phi )$ and $\chi ^{\prime }(\Phi ,K)$ are also
local functionals. This formula shows that the cohomologically non-trivial
solutions $\mathcal{G}^{\prime }(\phi )$ are just the gauge-invariant terms
constructed with the physical fields $\phi $ and their derivatives.

Assumption (\ref{autoassu}), instead of (\ref{galassu}), is actually
sufficient for the arguments that follow. Let $\{\mathcal{G}_{i}(\Phi )\}$
denote a basis for the non-trivial solutions $\mathcal{G}(\Phi )$ appearing
in (\ref{autoassu}). Extend the action $S_{c}$ of (\ref{basa}) replacing $%
\mathcal{S}(\Phi )$ with a linear combination $\mathcal{S}^{\prime }(\Phi )$
of all $\mathcal{G}_{i}(\Phi )$s, multiplied by independent parameters $%
\lambda _{i}$: 
\begin{equation}
S_{c}^{\prime }(\Phi ,K)=\mathcal{S}^{\prime }(\Phi )-\int R^{A}(\Phi
)K_{A},\qquad \mathcal{S}^{\prime }(\Phi )=\sum_{i}\lambda _{i}\mathcal{G}%
_{i}(\Phi ).  \label{exto}
\end{equation}
Since $(S_{c},\mathcal{G}_{i})=0$ and $(\mathcal{G}_{i},\mathcal{G}_{j})=0$
the extended action $S_{c}^{\prime }(\Phi ,K)$ still solves the master
equation $(S_{c}^{\prime },S_{c}^{\prime })=0$. From now on we drop the
primes in $S_{c}^{\prime }$ and $\mathcal{S}^{\prime }$ and assume that $%
S_{c}$ is the extended action.

Now, by assumption (\ref{autoassu}) we can decompose $\Gamma _{n\hspace{%
0.01in}\text{div}}^{(n+1)}$ as 
\[
\Gamma _{n\hspace{0.01in}\text{div}}^{(n+1)}=\sum_{i}(\Delta _{n+1}\lambda
_{i})\mathcal{G}_{i}(\Phi )+(S_{c},\chi _{n+1}), 
\]
where $\Delta _{n+1}\lambda _{i}$ are constants and $\chi _{n+1}$ is a local
functional. The divergences $\mathcal{G}_{i}(\Phi )$ are subtracted
redefining the parameters $\lambda _{i}$ as $\lambda _{i}^{\prime }=\lambda
_{i}-\Delta _{n+1}\lambda _{i}$, while the cohomologically trivial
divergences $(S_{c},\chi _{n+1})$ are subtracted by means of the canonical
transformation generated by 
\[
F_{n+1}(\Phi ,K^{\prime })=\int \Phi ^{A}K_{A}^{\prime }-\chi _{n+1}(\Phi
,K^{\prime }). 
\]
Indeed, 
\[
\Phi ^{\prime \hspace{0.01in}A}=\Phi ^{A}-\frac{\delta \chi _{n+1}}{\delta
K_{A}},\qquad K_{A}^{\prime }=K_{A}+\frac{\delta \chi _{n+1}}{\delta \Phi
^{A}},\qquad S_{n}(\Phi ^{\prime },K^{\prime })=S_{n}(\Phi ,K)-(S_{c},\chi
_{n+1}), 
\]
plus higher orders, therefore the action 
\[
S_{n+1}(\Phi ,K,\lambda )\equiv S_{n}(\Phi ^{\prime },K^{\prime },\lambda
^{\prime }) 
\]
is such that 
\[
S_{n+1}(\Phi ,K,\lambda )=S_{n}(\Phi ,K,\lambda )-\sum_{i}(\Delta
_{n+1}\lambda _{i})\mathcal{G}_{i}(\Phi )-(S_{c},\chi _{n+1})=S_{n}(\Phi
,K,\lambda )-\Gamma _{n\hspace{0.01in}\text{div}}^{(n+1)} 
\]
plus higher orders. Clearly, $S_{n+1}$ is the $(n+1)$-loop renormalized
action, because the functional $\Gamma _{n+1}$ is equal to $\Gamma
_{n}-\Gamma _{n\hspace{0.01in}\text{div}}^{(n+1)}$ plus higher orders.
Finally, $S_{n+1}$ also satisfies the master equation $(S_{n+1},S_{n+1})=0$,
since canonical transformations and parameter-redefinitions preserve the
antiparentheses. Thus, the inductive hypotheses are promoted to the order $%
n+1$. Iterating the argument, we find that $S_{\infty }$ is the renormalized
action to all orders and satisfies the master equation $(S_{\infty
},S_{\infty })=0$.

In ref. \cite{mastercan} the derivation just recalled was extended to the
master functional and the field-covariant proper formalism for gauge
theories. It was also shown that when the cohomological assumption (\ref
{autoassu}) holds, then it generalizes to an analogous cohomological
property for the proper formalism.

\section{Parameter-completion without cohomology}

\setcounter{equation}{0}

From now on we do not assume that the gauge algebra closes off shell, nor
that the symmetry satisfies particular cohomological properties, such as (%
\ref{autoassu}) and its generalizations. The only assumption we retain is
that the theory is manifestly free of gauge anomalies, namely $S_{c}$
satisfies the master equation 
\begin{equation}
(S_{c},S_{c})=0  \label{keymast}
\end{equation}
exactly at the regularized level. We show that renormalization itself allows
us to extend the classical action $S_{c}$ preserving the master equation
till the extended $S_{c}$ becomes parameter-complete.

We present our arguments in two steps. In this section, $i$) we introduce
enough redundant parameters so that all redefinitions of fields and sources
can actually be traded for redefinitions of the redundant parameters. In
section 5, $ii$) we remove those ad hoc parameters and take full advantage
of the possibility to make canonical transformations. Option $i$) is a
formal trick for intermediate derivations. Option $ii$) is the right way to
go to determine if our theory belongs to some special class with respect to
its renormalizability properties, for example it is finite, or
renormalizable with a finite number of parameters.

\paragraph{Raw renormalization algorithm \newline
}

Before deriving our main results, we need to recall a ``raw''
renormalization algorithm \cite{lavrov},
where divergences are subtracted
just as they come, without checking whether they can be reabsorbed into
parameter- and/or field-source-redefinitions. This construction allows us to
define a map that is crucial to build the extension map (\ref{extamap}).

As before, call $S_{n}$ and $\Gamma _{n}$ the action and the $\Gamma $%
-functional renormalized up to $n$ loops, with $S_{0}=S_{c}$. We allow the
starting action $S_{c}$ to be an expansion in $\hbar $, although we still
call it ``classical action''. Later we will appreciate why it is useful to
have an $\hbar $-dependent $S_{c}$. We denote the $\hbar \rightarrow 0$
limit of $S_{c}(\lambda ,\hbar )$ with $\bar{S}_{0}$. Since we use the
minimal subtraction scheme, $S_{n}=S_{c}+$poles and $S_{c}$ does not contain
evanescent terms equal to powers of $\varepsilon $ times finite local terms.

We inductively assume that $S_{n}$ satisfies the master equation up to
higher orders, namely 
\begin{equation}
(S_{n},S_{n})=\mathcal{O}(\hbar ^{n+1}).  \label{ostro}
\end{equation}
Applying the theorem (\ref{useidu}) we get the identity 
\begin{equation}
(\Gamma _{n},\Gamma _{n})=\langle (S_{n},S_{n})\rangle .  \label{useida}
\end{equation}
Using (\ref{ostro}), formula (\ref{useida}) gives $(\Gamma _{n},\Gamma _{n})=%
\mathcal{O}(\hbar ^{n+1})$. Now, $(S_{n},S_{n})$ is a local functional, and $%
\langle (S_{n},S_{n})\rangle $ is the functional that collects the
one-particle irreducible correlations functions containing one insertion of $%
(S_{n},S_{n})$. Because of (\ref{ostro}), the $\mathcal{O}(\hbar ^{n+1})$%
-contributions to $\langle (S_{n},S_{n})\rangle $ coincide with the $%
\mathcal{O}(\hbar ^{n+1})$-contributions to $(S_{n},S_{n})$. Moreover, since 
$S_{n}=S_{c}+$poles and $(S_{c},S_{c})=0$, we know that $(S_{n},S_{n})=$%
poles.

Call $\Gamma _{n\hspace{0.01in}\text{div}}^{(n+1)}$ the order-$(n+1)$
divergent part of $\Gamma _{n}$. By the theorem of locality of counterterms, 
$\Gamma _{n\hspace{0.01in}\text{div}}^{(n+1)}$ is a local functional. By the
observations just made, if we take the order-$(n+1)$ divergent part of (\ref
{useida}), we get 
\begin{equation}
(\bar{S}_{0},\Gamma _{n\hspace{0.01in}\text{div}}^{(n+1)})=\frac{1}{2}%
(S_{n},S_{n})+\mathcal{O}(\hbar ^{n+2}).  \label{cohoold}
\end{equation}
Now we define 
\begin{equation}
S_{n+1}=S_{n}-\Gamma _{n\hspace{0.01in}\text{div}}^{(n+1)}.  \label{buo}
\end{equation}
Clearly, $S_{n+1}$ is the $(n+1)$-loop renormalized action, since $\Gamma
_{n+1}=\Gamma _{n}-\Gamma _{n\hspace{0.01in}\text{div}}^{(n+1)}+\mathcal{O}%
(\hbar ^{n+2})$. Moreover, we still have $S_{\hspace{0.01in}n+1}=S_{c}+$%
poles, and, using (\ref{cohoold}) and (\ref{buo}), 
\[
(S_{n+1},S_{n+1})=\mathcal{O}(\hbar ^{n+2}), 
\]
which promotes the inductive assumption to $n+1$ loops. Iterating the
argument, we can construct the renormalized action $S_{\infty }$ and the
renormalized functional $\Gamma _{\infty }$, and prove that both satisfy
their master equations exactly.

Let us study $S_{\infty }$ more closely. In dimensional regularization the $%
L $-loop divergences are multiplied by 
\begin{equation}
\frac{\hbar ^{L}}{\varepsilon ^{n}},\qquad 1\leqslant n\leqslant L.
\label{diva}
\end{equation}
Thus, while $S_{c}$ depends on $\lambda $ and $\hbar $, $S_{\infty }$
depends on one additional quantity, which is $\hbar /\varepsilon $, and the
new dependence is (order-by-order) polynomial. Given a solution $S_{c}$ of
the master equation, the map 
\begin{equation}
S_{c}(\lambda ,\hbar )\rightarrow S_{\infty }(\lambda ,\hbar /\varepsilon
,\hbar )  \label{mappa}
\end{equation}
builds an \textit{extended solution} $S_{\infty }$ of the master equation,
such that the functional $\Gamma _{\infty }$ associated with $S_{\infty }$
is convergent. Since $S_{\infty }=S_{c}+$poles, we have $S_{\infty }(\lambda
,0,\hbar )=S_{c}(\lambda ,\hbar )$. We discover that renormalization knows
how to automatically extend the solutions of the master equation. This piece
of information is crucial for the arguments of this paper.

From the physical point of view, the raw subtraction algorithm is not the
final answer to the problem of renormalization, because when divergences are
subtracted just as they come, instead of by means of field-, parameter- and
source-redefinitions,\ renormalization-group invariance is lost. We cannot
define a bare action, because the renormalized action $S_{\infty }$ does not
contain enough independent constants to define all bare parameters we need.
To have RG invariance we must extend the classical action introducing new
independent parameters where appropriate.

\paragraph{Parameter-extension maps \newline
}

For the moment we adopt the option $i$) mentioned above and view
renormalization as a redefinition of parameters only, with no field/source
redefinitions. We prove that the classical action always admits an extension
that satisfies the master equation and contains enough independent
parameters to subtract all divergences by means of parameter-redefinitions.
The basic argument is that in case it is not so, we can use renormalization
to build an extended solution containing at least one additional independent
parameter. Iterating this procedure indefinitely, we end up with a
parameter-complete action, namely an action that can reabsorb its own
divergences redefining its own parameters.

We learned from (\ref{mappa}) that renormalization generates a new
perturbatively local solution $S_{\infty }$ of the master equation that
depends on one quantity ($\hbar /\varepsilon $) not contained in $S_{c}$.
Turning $\hbar /\varepsilon $ into an independent parameter $\lambda
^{\prime }$ we obtain the \textit{parameter-extension map} (often
abbreviated to \textit{extension map}) 
\begin{equation}
S_{c}(\lambda ,\hbar )\rightarrow S_{\infty }(\lambda ,\lambda ^{\prime
},\hbar ),  \label{mastermap}
\end{equation}
from a classical solution $S_{c}(\lambda ,\hbar )$ to the master equation,
depending on certain parameters $\lambda $, to an extended classical
solution $S_{\infty }(\lambda ,\lambda ^{\prime },\hbar )$, which can
(polynomially) depend on one additional parameter $\lambda ^{\prime }$. We
say that $S_{\infty }(\lambda ,\lambda ^{\prime },\hbar )$ is the
parameter extension of $S_{c}(\lambda ,\hbar )$.

Now, construct an \textit{extension chain} $\{S^{(0)},S^{(1)},S^{(2)}\cdots
\}$ of classical actions, all of which are solutions of the master equation,
where $S^{(0)}$ is the starting classical action $S_{c}(\lambda ,\hbar )$,
and $S^{(i)}$, $i>0$, is the parameter extension of $S^{(i-1)}$. Denote the
parameters contained in $S^{(i)}$ with $\lambda ^{(i)}$. We have $\{\lambda
^{(i)}\}\subset \{\lambda ^{(i+1)}\}$. Writing $\{\lambda ^{(i+1)}\}\equiv
\{\lambda ^{(i)},\lambda ^{\prime \hspace{0.01in}(i)}\}$, we also have $%
\left. S^{(i+1)}(\lambda ^{(i+1)},\hbar )\right| _{\lambda ^{\prime \hspace{%
0.01in}(i)}=0}=S^{(i)}(\lambda ^{(i)},\hbar )$.

If there exists an $i=I$ such that the parameter extension of $S^{(I)}$ is
stable up to parameter-redefinitions $\tilde{\lambda}$, namely such that 
\begin{equation}
S^{(I+1)}(\lambda ^{(I+1)},\hbar )=S^{(I)}(\tilde{\lambda}(\lambda
^{(I)},\lambda ^{\prime \hspace{0.01in}(I)},\hbar ),\hbar ),
\label{complete}
\end{equation}
we say that the extension chain closes. Then the action $S_{\subset }\equiv
S^{(I)}$ is parameter-complete. Indeed, by construction if we replace the
parameter $\lambda ^{\prime \hspace{0.01in}(I)}$ back with $\hbar
/\varepsilon $, formula (\ref{complete}) tells us that the action 
\[
S^{(I+1)}(\lambda ^{(I)},\hbar /\varepsilon ,\hbar )=S^{(I)}(\tilde{\lambda}%
(\lambda ^{(I)},\hbar /\varepsilon ,\hbar ),\hbar ) 
\]
is the renormalized action associated with the classical action $%
S^{(I)}(\lambda ^{(I)},\hbar )$, therefore $S_{\mathrm{B}}=S^{(I)}$ is the
bare action, 
\[
\lambda _{\mathrm{B}}=\tilde{\lambda}(\lambda ^{(I)},\hbar /\varepsilon
,\hbar ) 
\]
are the bare couplings and $\lambda ^{(I)}$ are the renormalized couplings.

In power-counting renormalizable theories closure is certainly achieved
after a finite number of steps, because the number of independent parameters
cannot exceed the number of monomials contained in the action. When the
theory is not power-counting renormalizable, instead, appropriate
truncations, discussed in the next section, are necessary to achieve closure
with a finite number of steps.

\paragraph{Reduced parameter-extension map \newline
}

We can always choose an $\hbar $-independent classical action $S_{c}$, or we
can replace $\hbar $ inside $S_{c}$ with a new independent parameter and add
it to the set of $\lambda $s. Then $S_{\infty }(\lambda ,\hbar /\varepsilon
,\hbar )$ depends on two parameters more than $S_{c}$, so the map (\ref
{mastermap}) extends the classical action by two parameters $\lambda
_{1}^{\prime }$ and $\lambda _{2}^{\prime }$, which replace $\hbar
/\varepsilon $ and $\hbar $, respectively. The extension chain can be
constructed as before, and the parameter-complete action satisfies 
\[
S^{(I+1)}(\lambda ^{(I+1)})=S^{(I)}(\tilde{\lambda}(\lambda ^{(I)},\lambda
^{\prime \hspace{0.01in}(I)})). 
\]

In this situation we can also construct a ``reduced'' extension map, which
can be useful for some purposes. It is obtained considering $S_{\infty
}(\lambda ,\hbar /\varepsilon ,0)$, that is to say keeping only the maximal
divergences of the renormalized action$\ S_{\infty }(\lambda ,\hbar
/\varepsilon ,\hbar )$. The reduced extension map 
\begin{equation}
S_{c}(\lambda )\rightarrow S_{\infty }(\lambda ,\lambda _{1}^{\prime },0)
\label{remap}
\end{equation}
is much easier to work out, since it is sufficient to compute the one-loop
divergent parts generated by $S_{c}(\lambda )$ and then use standard RG
techniques to resum the maximal divergences of diagrams with more loops.
Using (\ref{remap}) instead of (\ref{mastermap}), we can construct a reduced
extension chain and a reduced parameter-complete action $S_{r\subset }$. The
downside is that $S_{r\subset }$ is parameter-complete only with respect to
the maximal divergences. In some cases the action $S_{r\subset }$ may
coincide with the final answer $S_{\subset }$, because normally new
counterterms are generated already at one loop. However, we have no
guarantee that it is so (and it is quite easy to construct examples where it
is not so). A\ convenient strategy is to first construct the reduced action $%
S_{r\subset }$ and then check whether it is complete or not. It not, take $%
S_{r\subset }$ as the starting $S_{c}(\lambda )$ and build the complete
action $S_{\subset }$ using (\ref{mastermap}).

\section{Truncations}

\setcounter{equation}{0}

When we quantize a nonrenormalizable theory, or study composite fields of
high dimensions in any kind of theory, we have to define a consistent
perturbative expansion. In particular, we must truncate the classical action 
$S_{c}$ so that the truncated action $S_{cT}$ contains an arbitrarily large,
but finite, number $N$ of terms, sufficient for all practical needs. In the
previous section we constructed the parameter-complete action without paying
attention to this issue. Here we show how to truncate the theory and adapt
the construction of the previous section so that it involves a finite number
of operations for each $N$.

Denote the gauge coupling of minimum dimension with $\kappa $. We
parametrize the starting classical action $S_{c}(\Phi ,K,\kappa ,\zeta ,\xi
) $ as 
\begin{equation}
S_{c}(\Phi ,K,\kappa ,\zeta ,\xi )=\frac{1}{\kappa ^{2}}S_{c}^{\prime
}(\kappa \Phi ,\kappa K,\zeta ,\xi ),  \label{para}
\end{equation}
where $\xi $ are gauge-fixing parameters, $\zeta $ are any other parameters
and $S_{c}^{\prime }$ is polynomial in $\zeta $ and $\xi $. Moreover, each
field $\Phi $ has a dominant kinetic term 
\begin{equation}
\sim \frac{1}{2}\int \Phi \partial ^{n_{\Phi }}\Phi  \label{domina}
\end{equation}
normalized to one or multiplied by a dimensionless parameter.

Before proceeding let us explain the meaning of the parametrization (\ref
{para}). Consider for example Yang-Mills theory coupled to Einstein gravity.
The Yang-Mills action reads 
\begin{equation}
\frac{1}{4}\int \sqrt{g}g^{\mu \rho }g^{\nu \sigma }F_{\mu \nu }^{a}F_{\rho
\sigma }^{a}(A,g_{c}),  \label{bud}
\end{equation}
where $g_{c}$ is the gauge coupling. However, the gauge coupling $\kappa $
of minimum dimension is not $g_{c}$, but the Newton constant $\kappa _{N}$.
Thus (\ref{bud}) does not agree with (\ref{para}). The right way to
parametrize (\ref{bud}) is to define $g_{c}\equiv r_{+}\kappa _{N}$ and
rewrite (\ref{bud}) as 
\[
\frac{1}{4\kappa _{N}^{2}}\int \sqrt{g}g^{\mu \rho }g^{\nu \sigma }F_{\mu
\nu }^{a}F_{\rho \sigma }^{a}(\kappa _{N}A,r_{+}). 
\]
The action $S_{c}^{\prime }$ that we obtain is obviously polynomial in $%
r_{+} $.

When a theory contains superrenormalizable terms and massless fields Feynman
diagrams usually have infrared problems. To avoid this, we assume that if
superrenormalizable interactions are present, the fields are equipped with
appropriate quadratic terms that cure the infrared behaviors of diagrams.
For scalars and fermions we just need mass terms. For Yang-Mills theories in
three dimensions we need Chern-Simons terms. For higher-derivative gravities
in four dimensions, such as the theories with dominant quadratic terms 
\begin{equation}
\frac{1}{2\kappa _{N}^{2}}\int \sqrt{g}\left( \alpha R_{\mu \nu
}(D^{2})^{n}R^{\mu \nu }+\beta R(D^{2})^{n}R\right)  \label{hdgrav}
\end{equation}
where $\alpha $ and $\beta $ are dimensionless constants, $n\geqslant 0$ is
an integer and $D$ is the covariant derivative, we need either the Einstein
term or the cosmological term. Writing $g_{\mu \nu }=\bar{g}_{\mu \nu
}+\kappa _{N}\phi _{\mu \nu }$, where $\bar{g}_{\mu \nu }$ is some reference
metric, the fluctuation $\phi _{\mu \nu }$ has dimension $-n$. When $n>0$
the Newton constant $\kappa _{N}$ is a superrenormalizable parameter, so the
cosmological constant must be present anyway, because radiative corrections
generate it as a counterterm. Clearly, the theories (\ref{hdgrav}) are not
perturbatively unitary.

The gauge-fixing must be parametrized similarly. Let 
\[
S_{c\text{min}}(\Phi ,K,\kappa ,\zeta )=\frac{1}{\kappa ^{2}}S_{c\text{min}%
}^{\prime }(\kappa \Phi ,\kappa K,\zeta ) 
\]
denote the minimal solution of the master equation, namely $S_{c}(\Phi
,K,\kappa ,\zeta ,\xi )$ where antighosts $\bar{C}$, Lagrange multipliers $B$
and their sources $K_{\bar{C}}$, $K_{B}$ are set to zero. The simplest
extended solution of the master equation reads 
\[
S_{c\text{ext}}(\Phi ,K,\kappa ,\zeta )=S_{c\text{min}}(\Phi ,K,\kappa
,\zeta )-\int BK_{\bar{C}}=\frac{1}{\kappa ^{2}}S_{c\text{ext}}^{\prime
}(\kappa \Phi ,\kappa K,\zeta ), 
\]
and can be gauge-fixed using a gauge fermion $\Psi $ of the form 
\[
\Psi (\Phi ,K,\kappa ,\xi )=\frac{1}{\kappa ^{2}}\Psi ^{\prime }(\kappa \Phi
,\kappa K,\xi ), 
\]
where $\xi $ are gauge-fixing parameters and $\Psi ^{\prime }$ depends
polynomially on $\xi $. The $\Psi $-contributions that do gauge-fix are
actually contained in $\Psi (\Phi ,0,\kappa ,\xi )$, since the $K$-dependent
sector just describes a change of variables.

If the gauge algebra closes off shell we can choose an $S_{c\text{min}}$
linear in $K$. Taking a $K$-independent $\Psi $ the gauge-fixed solution of
the master equation reads 
\begin{equation}
S_{c}(\Phi ,K,\kappa ,\zeta ,\xi )=S_{c\text{ext}}+(S_{c\text{ext}},\Psi )=%
\frac{1}{\kappa ^{2}}S_{c}^{\prime }(\kappa \Phi ,\kappa K,\zeta ,\xi ).
\label{untrunc}
\end{equation}
For example, in Yang-Mills theory coupled with gravity a typical and simple
choice is 
\begin{eqnarray}
\Psi (\Phi ,K,\kappa ,\xi ) &=&\int \bar{C}^{\mu }\left( \eta ^{\hat{a}\nu
}\partial _{\nu }\phi _{\hat{a}\mu }+\xi _{G}\eta ^{\hat{a}\nu }\partial
_{\mu }\phi _{\hat{a}\nu }-\frac{\xi _{G}^{\prime }}{2}B_{\mu }\right) +\int 
\bar{C}^{\hat{a}\hat{b}}\phi _{\hat{a}\mu }\delta _{\hat{b}}^{\mu } 
\nonumber \\
&&+\int \bar{C}^{a}\left( \partial ^{\mu }A_{\mu }^{a}-\frac{\xi _{g}}{2}%
B^{a}\right) ,  \label{gf0}
\end{eqnarray}
where $\kappa \phi _{\hat{a}\mu }$ is the quantum fluctuation of the
vierbein $e_{\hat{a}\mu }$ around a given background (normally flat space), $%
\bar{C}^{\hat{a}\hat{b}}$ and $B_{\hat{a}\hat{b}}$ are the antighosts and
Lagrange multipliers of local Lorentz symmetry, $\hat{a}$, $\hat{b}$, $%
\ldots $ are indices of the Lorentz group and the indices of $\partial ^{\mu
}$, $\bar{C}^{\mu }$ and $B^{\mu }$ are raised and lowered with the flat
metric $\eta ^{\mu \nu }$.

More generally, if the gauge algebra closes only on shell we have to
gauge-fix the theory defining $S_{c}(\Phi ,K,\kappa ,\zeta ,\xi )$ as the
action obtained from $S_{c\text{ext}}$ applying the canonical transformation
generated by 
\begin{equation}
F(\Phi ,K^{\prime })=\int \Phi ^{A}K_{A}^{\prime }+\Psi (\Phi ,K^{\prime
},\kappa ,\xi ).  \label{cano}
\end{equation}
Clearly, $S_{c}$ is parametrized according to the structure (\ref{para}).

Now we can explain how the truncation works. We organize the set of
parameters $\lambda =\kappa ,\zeta ,\xi $ into four subsets $\bar{s}$, $%
s_{0} $, $s_{+}$ and $s_{-}$. The first subset $\bar{s}$ contains the
masses, the cosmological constant, and in general all parameters that enter
the propagator and are not treated perturbatively. For example, some of them
cannot be considered small because they cure infrared problems when
superrenormalizable interactions are present. We express each parameter
contained in $\bar{s}$ as a dimensionless constant of order one times $m^{d}$%
, where $d$ is its (non-negative) dimension in units of mass.

The second set $s_{0}$ contains the parameters of vanishing dimensions. We
write each of them as a constant of order one times a positive integer power
of some $\sigma \ll 1$. Then we have the subset $s_{+}$ of parameters that
have positive dimensions $d_{+}$ and are treated perturbatively, such as the
coefficients of superrenormalizable interactions. We write them as constants
of order one times $\Lambda _{+}^{d_{+}}$, where $\Lambda _{+}$ is some
scale. Finally, the forth subset $s_{-}$ contains the parameters of negative
dimensions $d_{-}$, which we write as constants of order one times $\Lambda
_{-}^{-d_{-}}$, where $\Lambda _{-}$ is some other scale. The forth subset
may include the coefficients of quadratic terms $\sim \phi \partial
^{n_{\phi }^{\prime }}\phi $ with $n_{\phi }^{\prime }>n_{\phi }$, which
have to be treated perturbatively, since we have established that the
dominant quadratic terms we perturb around are (\ref{domina}).

Feynman diagrams are multiplied by various factors, but their core integrals
depend only on the parameters of the subset $\bar{s}$ and external momenta.
Therefore, if we assume that $m$ and the
overall energy $E$ are of the same order, each field $\Phi $ of dimension $%
d_{\Phi }$ contributes to the amplitudes as a power $\sim E^{d_{\Phi }}\sim
m^{d_{\Phi }}$.

We assume that there exists a range of energies $E$ such that 
\begin{equation}
\Lambda _{+}\ll m\sim E\ll \Lambda _{-}  \label{range}
\end{equation}
and define the ratio 
\begin{equation}
\rho \sim \frac{E}{\Lambda _{-}}\sim \frac{\Lambda _{+}}{E}\ll 1.  \label{ro}
\end{equation}

In perturbatively unitary theories propagating fields have standard
dimensions in units of mass (1 for bosons and 3/2 for fermions, i.e. $%
n_{\Phi }=2$ and $n_{\Phi }=1$, respectively). When the theory is not
perturbatively unitary, like a higher-derivative theory, fields of
arbitrarily negative dimensions may be present. Including these theories is
useful to emphasize that our results are intrinsic properties of gauge
symmetry and renormalization, and do not depend on the particular model we
are working with.

The perturbative expansion is defined as the expansion in powers of $\rho $
and $\sigma $. The truncated actions are obtained neglecting the
contributions of orders $\rho ^{T^{\prime }}$ and $\sigma ^{T^{\prime }}$
with $T^{\prime }>T$, and denoted with $S_{cT}$, $S_{\infty T}$, $%
S_{T}^{(i)} $, $S_{\subset T}$, and so on. The $S_{cT}$-master equation must
hold within the truncation, which means $(S_{cT},S_{cT})=\mathcal{O}(\rho
^{T+1})+\mathcal{O}(\sigma ^{T+1})$. The other identities also hold up to $%
\mathcal{O}(\rho ^{T+1})$- and $\mathcal{O}(\sigma ^{T+1})$-corrections.

We show that $S_{cT}$ depends on a finite number of parameters and that
radiative corrections are compatible with the truncation. Let us first
assume that $S_{cT}$ is $\hbar $-independent. A generic term of $S_{cT}$ has
the structure 
\begin{equation}
(\kappa ^{2})^{L-1}\chi \partial ^{p}(\kappa \Phi )^{n_{\Phi }}(\kappa
K)^{n_{K}},  \label{structu2}
\end{equation}
with $L=0$, where $n_{\Phi }$ and $n_{K}$ are non-negative integer numbers
and $\chi $ is a product of parameters $\kappa $, $\zeta $ and $\xi $. The
structure (\ref{structu2}) for $L>0$ is the one of counterterms. Indeed,
since the action $S_{c}$ has an overall factor $1/\kappa ^{2}$ and a $\kappa 
$ is attached to each field and source, an $n$-leg vertex has at least a
factor $\kappa ^{n-2}$. Thus a diagram with $L$ loops, $I$ internal legs, $E$
external legs and $v_{jl}$ vertices with $n_{jl}=n_{\Phi j}+n_{Kl}$ legs,
where $n_{\Phi j}$ and $n_{Kl}$ are the numbers of $\Phi $- and $K$-legs,
respectively, is at least multiplied by a factor 
\[
\kappa ^{\sum_{jl}v_{jl}(n_{jl}-2)}=(\kappa ^{2})^{L-1}\kappa ^{E}, 
\]
in agreement with (\ref{structu2}). We have used the identities $L-I+V=1$
and $\sum_{jl}v_{jl}n_{jl}=2I+E$. We also derive an inequality that we need
below. Write $E=E_{\Phi }+E_{K}$, where $E_{\Phi }$ and $E_{K}$ are the
numbers of external $\Phi $- and $K$-legs of the diagram. Observing that $%
E_{K}=\sum_{jl}v_{jl}n_{Kl}$, we immediately get 
\[
E_{\Phi }+E_{K}=\sum_{jl}v_{jl}(n_{\Phi
j}+n_{Kl})-2I=E_{K}+\sum_{jl}v_{jl}(n_{\Phi j}-2)-2(L-1), 
\]
whence 
\begin{equation}
n_{\Phi j}\leqslant E_{\Phi }+2L.  \label{nif}
\end{equation}

Now, depending on whether the dimension $[\kappa ]$ of $\kappa $ is positive
or negative, we can write 
\[
\kappa \sim \Lambda _{\pm }^{[\kappa ]}=m^{[\kappa ]}\rho ^{|[\kappa ]|}. 
\]
If $[\kappa ]=0$ we write $\kappa =\sigma $. The terms (\ref{structu2})
belonging to the truncation must satisfy 
\begin{equation}
n_{\Phi },n_{K},2L\leqslant 2+\frac{T}{|[\kappa ]|^{e}},\qquad \chi _{\rho
}\leqslant T+2|[\kappa ]|^{e},  \label{bond1}
\end{equation}
with $e=1$ or $0$ for $[\kappa ]\neq 0$ and $[\kappa ]=0$, respectively,
while $\chi _{\rho }$ denotes the order of $\chi $.

The bounds (\ref{bond1}) are sufficient to show that the number of
derivatives $p$ must also be bounded from above. Indeed, in (\ref{structu2})
the product $\chi $ contributes with a factor $\Lambda _{+}^{u}\Lambda
_{-}^{-v}m^{w}$ for some non-negative $u,v,w$, and we must have $u+v\leqslant T+2|[\kappa ]|^{e}$. Now, given $n_{\Phi }$, $n_{K}$ and $L$, the dimension of $\chi \partial ^{p}$, which is equal to $u-v+w+p$, is also
given, but then the inequalities just found imply that $p$ must be bounded
from above. This proves that the truncation can contain only a finite number
of terms. We denote such number with $N(T)$.

Let us point out that the (\ref{bond1}) also implies that the number of
counterterms included in the truncation decreases when the order of
radiative corrections increases, and eventually drops to zero. Only a finite
number of Feynman diagrams contribute within the truncation, because the
number of loops and the number of vertices we can use are both bounded from
above.

When $S_{cT}$ depends on $\hbar $ we must assume that the parameters, or
product of parameters, that multiply the terms proportional to $\hbar ^{g}$
have an order $\sim \kappa ^{2g}$ higher than if they were tree-level. It is
possible to incorporate this assignment in formula (\ref{structu2})
replacing $L$ with $g$. Radiative corrections are also consistent.

Now that we know how to define an appropriately truncated action, we study
the extension map and make sure that it can be implemented with a finite
number of steps. There is a caveat, though. When we make the replacement $%
\hbar /\varepsilon \rightarrow \lambda ^{\prime }$ we lower the order of the
approximation. Indeed, by formula (\ref{structu2}), a factor $\kappa ^{2}$
appears at each loop, so defining the dimensionless constant $\tilde{\kappa}%
=\kappa m^{-[\kappa ]}$ we should consider $\lambda ^{\prime }$ as $\mathcal{%
O}(1/\tilde{\kappa}^{2})$, because it replaces a $\hbar /\varepsilon $.
However, it can be checked that if we really assume $\lambda ^{\prime }=%
\mathcal{O}(\tilde{\kappa}^{-2})$ then the replacement $\hbar /\varepsilon
\rightarrow \lambda ^{\prime }$ can generate infinitely many contributions
of the same order. To truncate such contributions we must assume $\lambda
^{\prime }=\mathcal{O}(\tilde{\kappa}^{\omega -2})$, with $\omega >0$. At
the same time, we must be sure that the radiative corrections to $\lambda
^{\prime }$, which are $\mathcal{O}(1)$ (see (\ref{run})) are smaller than $%
\lambda ^{\prime }$ itself, for which it is sufficient to assume $\omega <2$%
. Thus we take $0<\omega <2$.

Because the replacement $\hbar /\varepsilon \rightarrow \lambda ^{\prime }$
lowers the order of the approximation, it is not sufficient to truncate the
action to order $T$ to determine the extended action to order $T$. Instead,
we must truncate the classical action $S_{c}$ to some order $T_{0}$, to
determine the first extended action $S^{(1)}$ to some order $T_{1}$, so that
the second extended action $S^{(2)}$ is determined to some order $T_{2}$,
and so on, and guarantee that the final extended action $S_{\subset }$ is
determined to the desired order $T$. We want to show that this can be done
with a finite number of steps. In particular, $T_{0}(T)$ is finite.

Let us consider a single extension $S_{T_{i}}^{(i)}\rightarrow
S_{T_{i+1}}^{(i+1)}$. The renormalized action $S_{\infty
T_{i}}^{(i)}(\lambda ,\hbar /\varepsilon ,\hbar )$ contains terms (\ref
{structu2}) multiplied by 
\[
\left( \frac{\hbar }{\varepsilon }\right) ^{f}\hbar ^{g},\qquad f+g=L. 
\]
When we replace $\hbar /\varepsilon $ with $\lambda ^{\prime }$ we obtain
objects 
\[
\hbar ^{g}m^{(2-\omega )f[\kappa ]}\kappa ^{2g+\omega f-2}\chi \partial
^{p}(\kappa \Phi )^{n_{\Phi }}(\kappa K)^{n_{K}}. 
\]
We want to determine all terms of this type that fall within the truncation $%
T_{i+1}$. In particular, they must satisfy 
\[
n_{\Phi },n_{K},\omega f+2g\leqslant 2+\frac{T_{i+1}}{|[\kappa ]|^{e}}%
,\qquad \chi _{\rho }\leqslant T_{i+1}+2|[\kappa ]|^{e}, 
\]
therefore 
\[
L=f+g<f+\frac{2}{\omega }g\leqslant \frac{2}{\omega }+\frac{T_{i+1}}{\omega
|[\kappa ]|^{e}}. 
\]
Using (\ref{nif}) we see that all vertices $v$ participating in the diagrams
must satisfy 
\begin{equation}
n_{\Phi j}\leqslant \left( 2+\frac{T_{i+1}}{|[\kappa ]|^{e}}\right) \left( 1+%
\frac{2}{\omega }\right) ,\qquad n_{Kl}\leqslant 2+\frac{T_{i+1}}{|[\kappa
]|^{e}},\qquad \chi _{\rho }^{(v)}\leqslant T_{i+1}+2|[\kappa ]|^{e},
\label{bun}
\end{equation}
where $\chi _{\rho }^{(v)}$ is the order of the factor $\chi ^{(v)}$
appearing in the vertex. As before, given $n_{\Phi j}$ and $n_{Kl}$, the
number of derivatives $p_{v}$ that can appear in the vertex is also bounded
from above, because a large $p_{v}$ would raise the order of $\chi ^{(v)}$
arbitrarily. Thus, only a finite number of vertices can participate in the
diagrams that contribute to $S_{T_{i+1}}^{(i+1)}$. At this point, we
determine $T_{i}(T_{i+1})$ so that $S_{T_{i}}^{(i)}$ contains all such
vertices.

Recall that the parameter-complete action $S_{\subset T}$ we want to
determine contains a finite number of terms $N(T)$. Thus the extension chain 
$\{S_{T_{i}}^{(i)}(\lambda ^{(i)},\hbar )\}$ contains at most $N(T)$
elements, because each step adds at least one independent parameter, and
there cannot be more independent parameters than Lagrangian terms.
Consequently, $T_{0}(T)$ is finite, as we wished to prove. Thus, after a
finite number of operations we achieve closure within the truncation and
determine the parameter-complete action $S_{\subset T}$. That action is
equipped with all parameters that are necessary to renormalize divergences
by means of parameter-redefinitions, without using cohomological properties,
within the truncation.

Observe that choosing $\omega $ small the bounds (\ref{bun}) become larger,
which means that to determine the exended action more and more precisely as
a function of the new parameters $\lambda ^{\prime }$ we must work harder
and harder. These facts emphasize that our extension procedure is mainly a
theoretical tool. On the one side it is conceptually simple, on the other
side it appears to be prohibitive from the practical point of view, unless
ad hoc parameter-redefinitions and other tricks are found case-by-case to
reduce the effort.

\section{Parameter-completion and canonical transformations}

\setcounter{equation}{0}

So far we have used the approach $i$), where the extension algorithm is
applied after introducing redundant parameters to renormalize all kinds of
divergences, including those proportional to the field equations, by means
of parameter-redefinitions, instead of using both parameter-redefinitions
and canonical transformations of fields and sources. Now it is relatively
easy to explain how to proceed in the standard approach $ii$). We understand
that we are working with truncated actions where necessary, although we do
not make it explicit all the time.

The parameter-extension map is unchanged. Making the source- and
field-dependences explicit, we write (\ref{mastermap}) as 
\[
S_{c}(\Phi ,K,\lambda ,\hbar )\rightarrow S_{\infty }(\Phi ,K,\lambda
,\lambda ^{\prime },\hbar ). 
\]
The extension chain $\{S^{(0)},S^{(1)},S^{(2)}\cdots \}$ is obtained taking $%
S^{(0)}$ as the starting classical action $S_{c}(\lambda ,\hbar )$, and $%
S^{(i)}$, $i>0$, as the parameter extension of $S^{(i-1)}$. It is often
convenient to express each action $S^{(i)}$ in some specific field- and
source-variables, which we denote with $\Phi ^{(i)},K^{(i)}$. For example,
when the classical action is written in some standard form, we may want to
preserve that form throughout the extension process. This can be obtained
updating the field- and source-variables from $\{\Phi ^{(i-1)},K^{(i-1)}\}$
to $\{\Phi ^{(i)},K^{(i)}\}$ by means of canonical transformations. A common
option is to choose the ``essential'' form \cite{fieldcov,weinberg}, where
the dominant kinetic terms of the field equations (e.g. $\Box \phi $ and $%
\partial \!\!\!\slash\psi $ for bosons $\phi $ and fermions $\psi $ in
perturbatively unitary theories) are contained only in the dominant kinetic
terms of the action (up to total derivatives), and removed from every other
place by means of field redefinitions.

As before, the parameters contained in $S^{(i)}$ are denoted with $\lambda
^{(i)}$, and we have $\{\lambda ^{(i)}\}\subset \{\lambda
^{(i+1)}\}=\{\lambda ^{(i)},\lambda ^{\prime \hspace{0.01in}(i)}\}$. Now we
state that the chain closes if there exists an $i=I$ such that the
parameter extension of $S^{(I)}$ is stable in the sense that there exist
parameter-redefinitions $\tilde{\lambda}(\lambda ^{(I)},\lambda ^{\prime 
\hspace{0.01in}(I)},\hbar )$ and canonical transformations 
\[
\tilde{\Phi}(\Phi ,K,\lambda ^{(I)},\lambda ^{\prime \hspace{0.01in}%
(I)},\hbar ),\qquad \tilde{K}(\Phi ,K,\lambda ^{(I)},\lambda ^{\prime 
\hspace{0.01in}(I)},\hbar ), 
\]
such that 
\begin{equation}
S^{(I+1)}(\Phi ,K,\lambda ^{(I+1)},\hbar )=S^{(I)}(\tilde{\Phi},\tilde{K},%
\tilde{\lambda}(\lambda ^{(I)},\lambda ^{\prime \hspace{0.01in}(I)},\hbar
),\hbar ).  \label{closure2}
\end{equation}
When these operations are combined with the truncations explained above, the
extension chain closes after a finite number of manipulations. The action $%
S_{\subset }=S^{(I)}$ is then parameter-complete within the truncations.

Recapitulating, $S^{(I+1)}$ is the renormalized action and $S_{\mathrm{B}%
}=S^{(I)}$ is the bare action. Indeed, setting $\lambda ^{\prime \hspace{0.01in}(I)}
=\hbar/\varepsilon$ we obtain
\[
S^{(I+1)}(\Phi ,K,\lambda ^{(I)},\hbar /\varepsilon ,\hbar )=S_{\mathrm{B}}(\Phi _{\mathrm{B}},K_{\mathrm{B}},\lambda _{\mathrm{%
B}},\hbar ), 
\]
where
\[
\Phi _{\mathrm{B}}=\tilde{\Phi}(\Phi ,K,\lambda ^{(I)},\hbar /\varepsilon
,\hbar ),\qquad K_{\mathrm{B}}=\tilde{K}(\Phi ,K,\lambda ^{(I)},\hbar
/\varepsilon ,\hbar ),\qquad \lambda _{\mathrm{B}}=\tilde{\lambda}(\lambda
^{(I)},\hbar /\varepsilon ,\hbar ) 
\]
are the relations between bare and renormalized fields, sources and
parameters.

\section{Proper formalism and parameter-completion without cohomology}

\setcounter{equation}{0}

In the usual formalism of quantum field theory, based on the generating
functional $\Gamma $ of one-particle irreducible diagrams, any time the
canonical transformations are nonlinear, or contain field-dependent
source-trans\-form\-ations, they cannot be interpreted as true changes of
field variables in the functional integral, but only as replacements of
integrands \cite{fieldcov}. To overcome this issue, in ref. \cite{fieldcov}
we developed a field-covariant formalism for quantum field theory and in 
\cite{masterf} we introduced a new generating functional $\Omega $ of
one-particle irreducible diagrams, called ``master functional'', that
behaves as a scalar under arbitrary perturbative changes of field variables
(namely field redefinitions that can be expressed as local perturbative
series around the identity). The master functional supersedes the functional 
$\Gamma $, which does not transform in a simple way. In ref. \cite{mastercan}
the formalism was generalized to gauge theories.

The set of integrated fields is enlarged from $\Phi $ to a set of ``proper
fields'' $\Phi ,N$. Similarly, the set of sources $K$ is enlarged to the
``proper sources'' $K,H$. The extra fields $N^{I}$ are associated with local
composite fields $\mathcal{O}^{I}(\Phi )$, while the extra sources $H^{I}$
are associated with the $\mathcal{O}^{I}$-gauge transformations. The master
functional $\Omega $ and its classical action $S_{cN}$, called ``proper
action'', are functionals of the proper fields and the proper sources.

In ref. \cite{mastercan} it was shown that when cohomological properties
such as (\ref{autoassu}) hold in the usual formalism, they can be
generalized to the proper formalism and the master functional for gauge
theories. Doing so, a ``proper cohomology'' emerges, based on the \textit{%
squared antiparentheses} 
\begin{equation}
\left\lfloor X,Y\right\rfloor \equiv \int \left( \frac{\delta _{r}X}{\delta
\Phi ^{A}}\frac{\delta _{l}Y}{\delta K_{A}}+\frac{\delta _{r}X}{\delta N^{I}}%
\frac{\delta _{l}Y}{\delta H_{I}}-\frac{\delta _{r}X}{\delta K_{A}}\frac{%
\delta _{l}Y}{\delta \Phi ^{A}}-\frac{\delta _{r}X}{\delta H_{I}}\frac{%
\delta _{l}Y}{\delta N^{I}}\right)  \label{squart}
\end{equation}
between two functionals $X$ and $Y$ of $\Phi $, $K$, $N$ and $H$. The
squared antiparentheses satisfy identities analogous to the ones satisfied
by the usual antiparentheses, and can be used to extend the
Batalin-Vilkovisky formalism to the sector of composite fields.

Given a classical action $S_{c}(\Phi ,K)$ that satisfies (\ref{keymast}), it
is possible to construct a proper classical action $S_{cN}(\Phi ,K,N,H)$
that satisfies the proper master equation 
\begin{equation}
\lfloor S_{cN},S_{cN}\rfloor =0  \label{promast}
\end{equation}
and is such that the extra fields $N$ have ``propagator'' equal to one, and $%
S_{cN}=S_{c}$ at $H_{I}=0$, $\delta _{l}S_{cN}/\delta N^{I}=0$. The master
functional $\Omega $ collects the one-particle irreducible diagrams
generated by $S_{cN}$.

In \cite{mastercan} we used these tools to show that if (\ref{autoassu})
holds then $S_{cN}$ can be extended till it becomes parameter-complete. In
the proper approach a parameter-complete action contains enough independent
parameters so that all divergences can be removed by means of
parameter-redefinitions and ``proper'' canonical transformations, which are
special source-independent linear transformations of the proper fields $\Phi
,N$, combined with $\Phi $-$N$-independent source-transformations (see
formula (\ref{fcann})).

In the derivation of \cite{mastercan} cohomological properties played a
crucial role. On the other hand, in the previous sections we proved that
cohomological properties are not really essential for renormalization. In
this section we show that this fact remains true in the field-covariant
proper approach and construct a parameter-complete proper action $S_{\subset
N}$ without relying on cohomological assumptions.

As before, we assume that the starting classical action $S_{c}(\lambda
,\hbar )$ satisfies the master equation (\ref{keymast}) exactly at the
regularized level, which ensures that gauge anomalies are manifestly absent.
Then the proper classical action $S_{cN}$ also satisfies the proper master
equation (\ref{promast}) at the regularized level (see below for the proof).
To make the presentation more understandable, we first work with gauge
algebras that close off shell, and later generalize the results to gauge
algebras that close only on shell.

If the gauge algebra closes off shell we choose $S_{c}(\Phi ,K)$ of the form
(\ref{basa}). Then the starting proper action is 
\begin{equation}
S_{cN}(\Phi ,K,N,H)=S_{c}(\Phi ,K)-\int R_{\mathcal{O}}^{I}(\Phi )H_{I}+%
\frac{1}{2}\int \tilde{N}^{I}A_{IJ}\tilde{N}^{J}+\int \rho _{vI}\mathcal{N}%
^{v}(\tilde{N})\mathcal{O}_{\text{inv}}^{I}(\Phi ),  \label{gullo3}
\end{equation}
where $\tilde{N}^{I}=N^{I}-\mathcal{O}^{I}(\Phi )$, $A_{IJ}$ and $\rho _{vI}$
are constants, 
\[
R_{\mathcal{O}}^{I}(\Phi )\equiv \int R^{A}(\Phi )\frac{\delta _{l}\mathcal{O%
}^{I}}{\delta \Phi ^{A}} 
\]
are the gauge transformations of the composite fields $\mathcal{O}^{I}$, $%
\mathcal{O}_{\text{inv}}^{I}$ are the gauge-invariant composite fields and $%
\mathcal{N}^{v}(\tilde{N})$ denotes a basis of local monomials, at least
quadratic in $\tilde{N}$, constructed with $\tilde{N}$ and its derivatives.
The functional integral is over both $\Phi $ and $N$, and the ``improvement
term'' 
\begin{equation}
\frac{1}{2}\int N^{I}A_{IJ}N^{J},  \label{improvement}
\end{equation}
provides propagators for the fields $N$. All other quadratic terms coming
from the last contribution to (\ref{gullo3}) must be treated perturbatively
with respect to (\ref{improvement}). It is easy to check that (\ref{keymast}%
) indeed implies (\ref{promast}).

The generating functional of Green functions $Z$ and the generating
functional of connected Green functions $W$ are defined from 
\begin{equation}
Z(J,K,L,H)=\int [\mathrm{d}\Phi \hspace{0.01in}\mathrm{d}N]\exp \left(
-S_{cN}(\Phi ,K,N,H)+\int \Phi ^{A}J_{A}+\int N^{I}L_{I}\right) =\exp
W(J,K,L,H).  \label{zw}
\end{equation}
Since the $N$-propagator is equal to 1, the $N$-integral can be done exactly
in dimensional regularization, and amounts to make the Legendre transform of 
$S_{cN}$ with respect to $N^{I}$. Defining 
\[
-\tilde{S}(\Phi ,K,L,H)=-S_{cN}(\Phi ,K,N,H)+\int N^{I}L_{I},\qquad L_{I}=%
\frac{\delta _{l}S_{cN}}{\delta N^{I}}, 
\]
we get 
\[
\tilde{S}=\mathcal{S}(\Phi )-\int \left( R^{A}(\Phi )K_{A}+R_{\mathcal{O}%
}^{I}(\Phi )H_{I}\right) -\frac{1}{2}\int L_{I}(\tilde{A}^{-1})^{IJ}L_{J}-%
\int \tau _{vI}\mathcal{N}^{v}(L)\mathcal{O}_{\text{inv}}^{I}(\Phi )-\int 
\mathcal{O}^{I}L_{I}, 
\]
where the $\tau _{vI}$s are parameters equal to $\rho _{vI}$ plus
perturbative corrections and $\tilde{A}$ is the $A$-transpose. Thus we have 
\begin{equation}
Z=\int [\mathrm{d}\Phi ]\ \mathrm{\exp }\left( -\tilde{S}(\Phi ,K,L,H)+\int
\Phi ^{A}J_{A}\right) .  \label{nol}
\end{equation}
We see that $L_{I}$ play the role of sources coupled to the composite fields 
$\mathcal{O}^{I}$. Observe that the equations 
\[
L_{I}=\frac{\delta _{l}S_{cN}}{\delta N^{I}}=0, 
\]
which switch composite fields off, are solved by $\tilde{N}^{I}=0$.
Moreover, the conditions $H_{I}=$constants amount to a change of
gauge-fixing.

The master functional $\Omega $ is defined as the Legendre transform of $W$
with respect to both $\Phi $ and $L$, while $K$ and $H$ remain inert. We
have 
\begin{equation}
\Omega (\Phi ,K,N,H)=-W(J,K,L,H)+\int \left( \Phi
^{A}J_{A}+N^{I}L_{I}\right) ,  \label{hug}
\end{equation}
with 
\begin{eqnarray}
\Phi ^{A} &=&\frac{\delta _{r}W}{\delta J_{A}},\qquad N^{I}=\frac{\delta
_{r}W}{\delta L_{I}},\qquad \frac{\delta _{r}W}{\delta K_{A}}=-\frac{\delta
_{r}\Omega }{\delta K_{A}},  \nonumber \\
J_{A} &=&\frac{\delta _{l}\Omega }{\delta \Phi ^{A}},\qquad L_{I}=\frac{%
\delta _{l}\Omega }{\delta N^{I}},\qquad \frac{\delta _{r}W}{\delta H_{I}}=-%
\frac{\delta _{r}\Omega }{\delta H_{I}}.  \label{assur}
\end{eqnarray}
Since no confusion is expected to arise, we use the same names for the
proper fields $\Phi $, $N$ inside both $S_{N}$ and $\Omega $, while strictly
speaking the latter are averages of the former: $\langle \Phi _{S}\rangle
=\Phi _{\Omega }$, $\langle N_{S}\rangle =N_{\Omega }$. In the Legendre
transform (\ref{hug}) the improvement term (\ref{improvement}) must again be
treated as dominant with respect to all other quadratic terms involving the
fields $N^{I}$.

If the proper classical action $S_{cN}$ satisfies (\ref{promast}), then the
master functional $\Omega $ satisfies the proper master equation 
\[
\lfloor \Omega ,\Omega \rfloor =0. 
\]
The proof \cite{mastercan}, which we do not repeat here, follows from the
generalization of identity (\ref{useida}) to the proper formalism.

Changes of field variables and changes of gauge-fixings can be easily
implemented as ``proper'' canonical transformations, namely canonical
transformations for the proper fields and sources with generating functional
\begin{equation}
F(\Phi ,K^{\prime },N,H^{\prime })=\int (\Phi
^{A}+N^{I}b_{I}^{A})K_{A}^{\prime }+\int N^{I}z_{I}^{J}(H_{J}^{\prime }-\xi
_{J}),  \label{fcann}
\end{equation}
where $b_{I}^{A}$, $z_{I}^{J}$ and $\xi _{I}$ are constants, which can be
both $c$-numbers and Grassmann variables. More explicitly, a proper
transformation reads 
\begin{eqnarray}
\Phi ^{A\hspace{0.01in}\prime } &=&\frac{\delta F}{\delta K_{A}^{\prime }}%
=\Phi ^{A}+N^{I}b_{I}^{A},\qquad K_{A}=\frac{\delta F}{\delta \Phi ^{A}}%
=K_{A}^{\prime },  \nonumber \\
N^{I\hspace{0.01in}\prime } &=&\frac{\delta F}{\delta H_{I}^{\prime }}%
=N^{J}z_{J}^{I},\qquad H_{I}=\frac{\delta F}{\delta N^{I}}%
=z_{I}^{J}(H_{J}^{\prime }-\xi _{J})+b_{I}^{A}K_{A}^{\prime }.  \label{buah2}
\end{eqnarray}

Let us briefly describe how the field redefinitions contained in (\ref{buah2}%
) work. If we write 
\begin{equation}
\Phi ^{A\hspace{0.01in}\prime }=\Phi ^{A}+\mathcal{O}^{I}(\Phi )b_{I}^{A}+%
\tilde{N}^{I}b_{I}^{A},  \label{bisu}
\end{equation}
when we set $\tilde{N}^{I}=K=0$, $H=$constant, to switch off the sectors of
composite fields and gauge transformations, (\ref{bisu}) becomes $\Phi ^{A%
\hspace{0.01in}\prime }=\Phi ^{A}+\mathcal{O}^{I}(\Phi )b_{I}^{A}$, which is
the expansion of the most general local perturbative change of field
variables. However, the conditions $\tilde{N}^{I}=0$ switch off the
composite-field sector only before the transformation. Indeed, due to the
term $\tilde{N}^{I}b_{I}^{A}$ in (\ref{bisu}) after the transformation the
solutions of $L_{I}^{\prime }=\delta _{l}S_{N}^{\prime }/\delta N^{I\prime 
\hspace{0.01in}}=0$ at $K=0$, $H=$constant, are no longer $\tilde{N}^{I}=0$,
but some new $\tilde{N}^{I\hspace{0.01in}\prime }=0$. Working out $\tilde{N}%
^{I\hspace{0.01in}\prime }$ it is found that at $\tilde{N}^{I\hspace{0.01in}%
\prime }=K=0$, $H=$constant the effective change of variables is corrected
by $\mathcal{O}(b^{2})$-terms and finally reads 
\[
\Phi ^{A\hspace{0.01in}\prime }=\Phi ^{A}+\mathcal{O}^{I}(\Phi )\tilde{b}%
_{I}^{A}, 
\]
where $\tilde{b}_{I}^{A}=b_{I}^{A}+\mathcal{O}(b^{2})$ is some calculable
power series in $b$. More details can be found in refs. \cite
{mastercan,masterf}.

Now we generalize the arguments of the previous sections to the proper
formalism for gauge theories. The raw renormalization algorithm is
immediately generalized replacing the classical action $S_{c}$ with the
classical proper action $S_{cN}$, the antiparentheses with the squared
antiparentheses, the master equation with the proper master equation and the 
$\Gamma $-functional with the master functional $\Omega $. We do not repeat
the derivation here, because it was already given in section 5 of ref. \cite
{mastercan}. Calling $S_{N\hspace{0.01in}n}$ and $\Omega _{n}$ the proper
action and the master functional renormalized up to $n$ loops (with $S_{N%
\hspace{0.01in}0}=S_{cN}$), we subtract the order-$(n+1)$ divergent part $%
\Omega _{n\hspace{0.01in}\text{div}}^{(n+1)}$ of $\Omega _{n}$ (in the
minimal subtraction scheme) and define 
\begin{equation}
S_{N\hspace{0.01in}n+1}=S_{N\hspace{0.01in}n}-\Omega _{n\hspace{0.01in}\text{%
div}}^{(n+1)}.  \label{buom}
\end{equation}
Iterating this operation we construct the renormalized proper action $S_{N%
\hspace{0.01in}\infty }$ and prove that it satisfies the master equation 
\[
\lfloor S_{N\hspace{0.01in}\infty },S_{N\hspace{0.01in}\infty }\rfloor =0. 
\]
Now, if $\lambda $ denotes the parameters contained in $S_{cN}$, the map 
\begin{equation}
S_{cN}(\Phi ,K,N,H,\lambda ,\hbar )\rightarrow S_{N\hspace{0.01in}\infty
}(\Phi ,K,N,H,\lambda ,\hbar /\varepsilon ,\hbar )  \label{pap}
\end{equation}
sends a solution of the proper master equation into an extended solution.
Thus we can define the parameter-extension map 
\begin{equation}
S_{cN}(\Phi ,K,N,H,\lambda ,\hbar )\rightarrow S_{N\hspace{0.01in}\infty
}(\Phi ,K,N,H,\lambda ,\lambda ^{\prime },\hbar ),  \label{pappa}
\end{equation}
and build an extension chain $\{S_{N}^{(0)},S_{N}^{(1)},S_{N}^{(2)}\cdots \}$
of proper classical actions, where $S_{N}^{(0)}=S_{cN}$, and $S_{N}^{(i)}$, $%
i>0$, is the parameter extension of $S_{N}^{(i-1)}$. We state that the
extension chain closes if there exists an $i=I$ such that 
\begin{equation}
S_{N}^{(I+1)}(\Phi ,K,N,H,\lambda ^{(I+1)},\hbar )=S_{N}^{(I)}(\tilde{\Phi},%
\tilde{K},\tilde{N},\tilde{H},\tilde{\lambda}(\lambda ^{(I)},\lambda
^{\prime \hspace{0.01in}(I)},\hbar ),\hbar ),  \label{closure}
\end{equation}
where the tilded proper fields and sources are related to the untilded ones
by a proper canonical transformation that depends on $\lambda ^{(I)}$, $\lambda ^{\prime \hspace{0.01in}(I)}$ and $\hbar$. Within the truncations, the extension
chain closes after a finite number of operations, and $S^{(I)}$ identifies
the parameter-complete proper action $S_{\subset N}$.

Although we have assumed that the gauge algebra closes off shell, so far,
this assumption did not enter the key-steps of our arguments. Its main
purpose was to let us use the simple and explicit form (\ref{gullo3}) of the
starting proper action $S_{cN}$. Relaxing the assumption of off-shell
closure, we can start from any classical action $S_{c}(\Phi ,K)$ that
satisfies (\ref{keymast}). Then we pick a basis $\{\mathcal{O}^{I}(\Phi )\}$
of local composite fields and make a canonical transformation with
generating function 
\[
F_{c}(\Phi ,K^{\prime })=\int \Phi ^{A}K_{A}^{\prime }-\int \mathcal{O}%
^{I}(\Phi )H_{I}. 
\]
The transformed action reads 
\begin{equation}
\tilde{S}_{c}(\Phi ,K,H)=S_{c}(\Phi ,K+\int \frac{\delta _{l}\mathcal{O}^{I}%
}{\delta \Phi }H_{I})  \label{sexu}
\end{equation}
and obviously satisfies $(\tilde{S}_{c},\tilde{S}_{c})=0$. Define the proper
action 
\begin{equation}
S_{cN}(\Phi ,K,N,H)=\tilde{S}_{c}(\Phi ,K,H)+\frac{1}{2}\int \tilde{N}%
^{I}A_{IJ}\tilde{N}^{J},  \label{sext}
\end{equation}
where $\tilde{N}^{I}=N^{I}-\mathcal{O}^{I}$. It is easy to prove that $%
S_{cN} $ satisfies the proper master equation (\ref{promast}). First observe
that $\lfloor \tilde{S}_{c},\tilde{S}_{c}\rfloor $ $=(\tilde{S}_{c},\tilde{S}%
_{c})=0$, therefore 
\[
\lfloor S_{cN},S_{cN}\rfloor =\lfloor \tilde{S}_{c},\int \tilde{N}A\tilde{N}%
\rfloor . 
\]
Next, working out the squared antiparentheses explicitly, formulas (\ref
{sexu}) and (\ref{sext}) give 
\[
\lfloor \tilde{S}_{c},\tilde{N}^{I}\rfloor =\int \frac{\delta _{r}\tilde{S}%
_{c}}{\delta K_{A}}\frac{\delta _{l}\mathcal{O}^{I}}{\delta \Phi ^{A}}-\frac{%
\delta _{r}\tilde{S}_{c}}{\delta H_{I}}=0, 
\]
whence $\lfloor S_{cN},S_{cN}\rfloor =0$ immediately follows.

The generating functionals are defined as in (\ref{zw}) and (\ref{hug}).
Integrating over the extra fields $N$ we obtain (\ref{nol}) with 
\[
\tilde{S}(\Phi ,K,H,L)=\tilde{S}_{c}(\Phi ,K,H)-\frac{1}{2}\int L_{I}(\tilde{%
A}^{-1})^{IJ}L_{J}-\int \mathcal{O}^{I}L_{I}. 
\]
All other derivations are identical to the ones given before, since the
explicit forms of $S_{c}$ and $S_{cN}$ play no role in those. Again, we have
the parameter-extension map (\ref{mappa}), which allows us to construct the
extension chain. We define closure by formula (\ref{closure}), where the
canonical transformation must be proper. The truncations ensure that the
chain closes after a finite number $I$ of steps. The parameter-complete
proper action $S_{\subset N}=S^{(I)}$ can thus be worked out with a finite
number of calculations, and is such that its divergences can be subtracted
redefining parameters and making proper canonical transformations.

\section{Remarks}

\setcounter{equation}{0}

In this section we collect a few observations that can make us better
appreciate some properties of the constructions presented so far.

Every time we extend the solution with the operations described above, we
introduce a new parameter $\lambda ^{\prime }$, obtained replacing $\hbar
/\varepsilon $. Since the factors $\hbar /\varepsilon $ multiply powers of
other parameters $\lambda $ coming from Feynman diagrams, the new parameter $%
\lambda ^{\prime }$ also multiplies various powers of $\lambda $. Thus the
extended actions are parametrized in non-standard ways.

Call the coefficient of a Lagrangian monomial constructed with the fields,
the sources and their derivatives, \textit{parameter-singlet} if it is made
of a single parameter. Call it \textit{parameter-product} if it is made of a
product of parameters, with various (possibly negative) exponents. It may be
convenient to organize the action so that, proceeding order-by-order along
with the truncations defined previously, the first time a new parameter
appears it appears as a parameter-singlet. To achieve this goal, the first
time we find a Lagrangian term multiplied by a new independent coefficient
equal to a parameter-product, say $\lambda \lambda ^{\prime }$, we redefine
that coefficient as a new parameter-singlet $\alpha $, and re-express $%
\alpha $ everywhere else in terms of $\lambda $ and $\lambda ^{\prime }$. If
the coefficient is a sum of more parameter-products, to avoid complicated
functions of parameters we call $\alpha $ one parameter-product of the sum,
randomly chosen. When we do these operations, we very likely generate
negative powers of parameters, which makes the new parametrization also
non-standard. Ultimately, the original parametrization $S_{\subset }(\Phi
,K,\lambda ,\hbar )$ may be the most convenient one, because at least it
guarantees that all parameters appear polynomially.

Let us emphasize that if we search for the most general solution $%
S_{\sqsubset }$ of the master equation (see next section) we get coupled
quadratic equations (\ref{auto}) that can lead to even more involved
non-standard parametrizations, when cohomological properties do not hold.

Another remark concerns possible modifications of the gauge symmetry. When
the gauge algebra closes off shell and the cohomological property (\ref
{autoassu}) holds, the symmetry transformations are not affected by
radiative corrections in any observable way, and the renormalized action is
really equivalent to the starting classical action $S_{c}$, extended as
shown in (\ref{exto}), up to canonical transformations and
parameter-redefinitions. Instead, when (\ref{autoassu}) does not hold there
is no obvious reason why the gauge symmetry should remain the same after
renormalization. Radiative corrections can modify it in a physically
observable way. Moreover, even when the starting gauge symmetry, encoded in $%
S_{c}$, closes off shell, the final one, encoded in $S_{\subset }$ may close
only on shell.

We do not have examples of non-trivial parameter extensions induced by
renormalization, since the theories we normally deal with do satisfy (\ref
{autoassu}). Nevertheless, to clarify the remark just made it may be helpful
to think of an interacting gauge theory as a parameter extension of its
free-field limit. For example, switching off antighosts $\bar{C}$ and
Lagrange multipliers $B$, the solution of the master equation of an Abelian
gauge theory is just 
\[
S_{\text{Ab}}=\frac{1}{4}\int (\partial _{\mu }A_{\nu }^{a}-\partial _{\nu
}A_{\mu }^{a})^{2}+\int K_{\mu }\partial _{\mu }C. 
\]
If we assume that the number of photons is appropriate, we can write the
action $S_{\text{nAb}}$ of non-Abelian Yang-Mills theory as 
\[
S_{\text{nAb}}=S_{\text{Ab}}+\omega (A)-\int (\Delta R^{A})K_{A}, 
\]
where 
\begin{eqnarray*}
\omega (A) &=&gf^{abc}\int (\partial _{\mu }A_{\nu }^{a})A_{\mu }^{b}A_{\nu
}^{c}+\frac{g^{2}}{4}\int (f^{abc}A_{\mu }^{b}A_{\nu }^{c})^{2}, \\
-\int (\Delta R^{A})K_{A} &=&g\int \left[ K_{\mu }^{a}f^{abc}A_{\mu
}^{b}C^{c}+\frac{1}{2}K_{C}^{a}f^{abc}C^{b}C^{c}\right] .
\end{eqnarray*}
The first line is a one-parameter extension of the action, while the second
line encodes the corresponding extensions of the symmetry transformations.

This observation illustrates what may happen in more involved theories,
where new interactions may be created by the extension. In that case the
symmetry transformations may be modified accordingly, to preserve the master
equation.

\section{Search for the most general solution of the master equation}

\setcounter{equation}{0}

In the previous sections we proved that a manifestly anomaly-free gauge
theory always admits a parameter-complete classical action $S_{\subset }$.
A\ related issue that remains to be addressed is the search for the most
general solution $S_{\sqsubset }$ of the master equation. In this section we
make some remarks about this topic and compare the properties of the actions 
$S_{\sqsubset }$ and $S_{\subset }$.

Certainly $S_{\sqsubset }$ is parameter-complete, because it includes $%
S_{\subset }$. Another way to prove the parameter-completeness of $%
S_{\sqsubset }$ is to take $S_{\sqsubset }$ as the classical action and use
the raw renormalization algorithm to work out the renormalized action $%
S_{\infty \sqsubset }$. Since $S_{\sqsubset }$ is the most general solution
of the master equation, $S_{\infty \sqsubset }$ must be related to $%
S_{\sqsubset }$ by means of parameter-redefinitions and canonical
transformations. Observe that both these arguments do not make use of
cohomological properties.

Let us now see how $S_{\sqsubset }$ can be built. Start from any solution $%
S_{c}(\Phi ,K,\lambda )$ of the master equation and let $\{\mathcal{L}%
^{i}(\Phi ,K,\lambda )\}$ denote a basis of Lagrangian terms $\mathcal{L}%
^{i} $ constructed with the fields, the sources and their derivatives. For
simplicity, we can take a basis made of monomials. Composite fields $%
\mathcal{O}^{I}$ can be included also, coupled to external sources $L_{I}$.
The most general extension of the starting classical action $S_{c}$ can be
parametrized as 
\begin{equation}
S_{\sqsubset }(\Phi ,K,\lambda ,\tau )=S_{c}(\Phi ,K,\lambda )+\tau
_{i}\Delta ^{i}(\Phi ,K,\lambda ),\qquad \Delta ^{i}=\int \mathcal{L}^{i},
\label{extr}
\end{equation}
where the sum over $i$ is understood and the $\tau _{i}$s are constants.
Requiring that $S_{\sqsubset }$ solve the master equation $(S_{\sqsubset
},S_{\sqsubset })=0$ we obtain the condition 
\[
2\tau _{i}(S_{c},\Delta ^{i})+\tau _{i}\tau _{j}(\Delta ^{i},\Delta ^{j})=0 
\]
for the constants $\tau $. The objects $(\Delta ^{i},\Delta ^{j})$ are local
functionals, equal to the integrals of local composite fields of dimension 5
and ghost number one. Let $\Delta _{5}^{a}$ denote a basis of such
functionals. Then there exist constants $C_{a}^{ij}$ such that 
\[
(\Delta ^{i},\Delta ^{j})=C_{a}^{ij}\Delta _{5}^{a}, 
\]
where the sum over $a$ is understood. Expanding $S_{c}$ as $\sigma
_{i}\Delta ^{i}$, where $\sigma _{i}$ are known constants, the equations we
must solve can be written as 
\begin{equation}
(\sigma _{i}+\tau _{i})C_{a}^{ij}(\sigma _{j}+\tau _{j})=0,\qquad \sigma
_{i}C_{a}^{ij}\sigma _{j}=0.  \label{auto}
\end{equation}
The latter is a constraint on the $\sigma _{i}$s, following from $%
(S_{c},S_{c})=0$. Clearly, $\tau _{i}=h\sigma _{i}$, where $h$ is an overall
constant, are solutions, but they just give $S_{\sqsubset }=(1+h)S_{c}$.

If we do not assume cohomological properties such as (\ref{autoassu}), the
problem remains quadratic. The coupled quadratic algebraic equations (\ref
{auto}) can be very difficult to solve, and it is not even evident how to
solve them perturbatively. Thus, the search for $S_{\sqsubset }$ might not
be a practically viable strategy.

If we ignore the difficulties to build $S_{\sqsubset }$ and just assume that 
$S_{\sqsubset }$ is known, we can investigate its cohomological properties.
Denote the $S_{\sqsubset }$-independent parameters with $\rho _{i}$.
Differentiating $(S_{\sqsubset },S_{\sqsubset })=0$ with respect to $\rho $,
we find 
\begin{equation}
\left( S_{\sqsubset },\frac{\partial S_{\sqsubset }}{\partial \rho _{i}}%
\right) =0.  \label{acco}
\end{equation}
If $S_{\sqsubset }$ depends on $\hbar $ what is actually important is the $%
\hbar \rightarrow 0$-limit of this equation, which reads 
\begin{equation}
\left( \bar{S}_{0},\frac{\partial \bar{S}_{0}}{\partial \rho _{i}}\right) =0,
\label{acco2}
\end{equation}
$\bar{S}_{0}$ being the $\hbar \rightarrow 0$-limit of $S_{\sqsubset }$. We
see that the $\rho $-derivatives of $\bar{S}_{0}$ are solutions of the
cohomological problem (\ref{chopr}) with $S_{c}$ replaced by $\bar{S}_{0}$.
However, there is no guarantee that all non-trivial solutions of that
problem are contained in the set $\{\partial \bar{S}_{0}/\partial \rho
_{i}\} $.

Take $S_{\sqsubset }$ as the starting classical action. As usual, call $%
S_{n} $ the action renormalized up to $n$ loops, with $S_{0}=S_{\sqsubset }$%
, and assume that it satisfies the master equation $(S_{n},S_{n})=0$
exactly. Then the $(n+1)$-loop divergences $\Gamma _{n\hspace{0.01in}\text{%
div}}^{(n+1)}$ satisfy the cohomological problem $(\bar{S}_{0},\Gamma _{n%
\hspace{0.01in}\text{div}}^{(n+1)})=0$. If we want to remove $\Gamma _{n%
\hspace{0.01in}\text{div}}^{(n+1)}$ redefining the $\rho $s and making
canonical transformations, we need to know either that $a$) the set $%
\{\partial \bar{S}_{0}/\partial \rho _{i}\}$ contains all the non-trivial
solutions of the cohomological problem, or anyway $b$) it contains the
solutions generated by renormalization, namely there exist constants $%
\Delta _{n+1}\rho _{i}$ and local functionals $\chi _{n+1}$ such that 
\[
\Gamma _{n\hspace{0.01in}\text{div}}^{(n+1)}=\sum_{i}\Delta _{n+1}\rho _{i}%
\frac{\partial \bar{S}_{0}}{\partial \rho _{i}}+(\bar{S}_{0},\chi _{n+1}). 
\]
However, it is not enough to know that $S_{\sqsubset }$ is the most general
solution of the master equation to prove $a$). In principle, there might be
solutions of the cohomological problem that cannot be embedded into any
extension of the classical action. As far as statement $b$) is concerned,
the parameter-completeness of $S_{\sqsubset }$ proves that it does hold,
but, again, this argument does not use cohomological properties.

To conclude, instead of trying to solve the problem in a purely algebraic
way, it is more convenient to let renormalization build the extended action,
as we have done in this paper. Then we discover that a parameter-complete
solution $S_{\subset }$ always exists and obeys property $b$) (with $\bar{S}%
_{0}$ replaced by the $\hbar \rightarrow 0$-limit of $S_{\subset }$), but
not necessarily property $a$). The most general solution $S_{\sqsubset }$ of
the master equation is also parameter-complete, and obeys $b$). The action $%
S_{\sqsubset }$ may coincide with $S_{\subset }$ or be more general than $%
S_{\subset }$. In either case, statement $a$) is not guaranteed to hold.
Ultimately, only cohomological theorems can ensure a property as strong as $%
a $), but renormalization does not need that much.

\section{Conclusions}

\setcounter{equation}{0}

In this paper we have shown that in a manifestly anomaly-free gauge theory
it is always possible to extend the classical action $S_{c}$ into a
parameter-complete action $S_{\subset }$ that satisfies the master equation
and is able to remove all divergences redefining its own parameters and
making canonical transformations. The construction was also extended to the
master functional and the field-covariant proper formalism for gauge
theories, where renormalization works by means of parameter-redefinitions
and proper canonical transformations. Such canonical transformations are
true changes of variables in functional integrals and generating
functionals, rather than mere replacements of integrands.

The compatibility between gauge symmetry and renormalization is encoded in
an intrinsically quadratic problem, because the master equation is quadratic
in the action. The main virtue of our algorithm is that it solves the
quadratic problem even when it is not possible to reduce it to a much
simpler, linear (cohomological) problem. Cohomological properties can
linearize the quadratic problem, but their proofs must be worked out
case-by-case and normally demand a remarkable effort. Sufficiently powerful
cohomological theorems might not hold in the theory we are insterested in.
Even when they hold, we might just not want to use them. It is interesting
to know that, without assuming cohomological properties, whenever gauge
anomalies are manifestly absent the classical action can be iteratively
extended till it becomes parameter-complete. In other words, quantum field
theory, renormalization and renormalization-group invariance are
intrinsically compatible with gauge symmetry.

At the practical level, we start from any classical action $S_{c}(\Phi ,K)$.
In case we want to work with the proper formalism and the master functional,
we construct the proper action (\ref{sext}). Then we calculate the
renormalization of the theory, which allows us to define the maps (\ref
{mappa}) or (\ref{pap}). At this point we discover that renormalization is
able to build an extended classical action, which still satisfies the master
equation, but contains a new independent parameter. Taking advantage of this
fact and iterating the extension till it closes, we end up with the
parameter-complete action, $S_{\subset }$ or $S_{N\subset }$, which
satisfies the master equation and is able to renormalize all divergences by
means of parameter-redefinitions and (proper) canonical transformations. The
renormalization algorithm defined by this procedure is conceptually simpler
than any previously known one.

The results of this paper lead us to conjecture that if the theory is
potentially plagued with gauge anomalies, but admits anomaly cancellation at
one loop, it is always possible to find a parameter-complete extension of
the classical action such that the functionals $\Gamma $ and $\Omega $
satisfy their master equations exactly in the physical limit. However, we
have to leave the investigation of this issue to a separate work.

\vskip 12truept \noindent {\large \textbf{Acknowledgements}}

\vskip 2truept

The investigation of this paper was carried out as part of a program to
complete the book \cite{webbook}, which will be available at %
\href{http://renormalization.com}{\texttt{Renormalization.com}} once
completed. I\ thank the Perimeter Institute, Waterloo, Ontario, Canada, for
hospitality during the first stage of this work and the Physics Department
of Fudan University, Shanghai, for hospitality during the final stage of
this work.

\end{document}